\documentclass[twocolumn,showkeys,aps,amssymb,floatfix,prd]{revtex4}
\usepackage{bm,graphicx}

\newcommand\beq{\begin{equation}}
\newcommand\beqa{\begin{eqnarray}}
\newcommand\beqan{\begin{eqnarray*}}
\newcommand\eeq{\end{equation}}
\newcommand\eeqa{\end{eqnarray}}
\newcommand\eeqan{\end{eqnarray*}}

\newcommand\Mbh{M_\bullet}
\newcommand\gravr{{\sf m}_\bullet}
\newcommand\thbh{\theta_\bullet}
\newcommand\vthE{\vartheta_E}
\newcommand\vep{\varepsilon}
\newcommand\vth{\vartheta}
\newcommand\vthbh{\vartheta_\bullet}
\newcommand\cb{{\cal B}}
\newcommand\RN{Reissner-Nordstr\"om}
\newcommand\order[2]{\mathcal{O}\left({#1}\right)^{#2}}

\newcommand\reffig[1]{Figure~\ref{fig:#1}}
\newcommand\reffigs[2]{Figures~\ref{fig:#1} and \ref{fig:#2}}

\begin{document}

\title{Formalism for Testing Theories of Gravity Using \\
Lensing by Compact Objects. I:  Static, Spherically Symmetric Case}

\author{Charles R.\ Keeton}
\affiliation{Department of Physics \& Astronomy, Rutgers University,
  136 Frelinghuysen Road, Piscataway, NJ 08854;\\
  {\tt keeton@physics.rutgers.edu}}

\author{A.\ O.\ Petters}
\affiliation{Departments of Mathematics and Physics,
  Duke University,\\
  Science Drive,\\ Durham, NC 27708-0320;\\
  {\tt petters@math.duke.edu}}

%\date{Accepted in PRD October 21, 2005}

\begin{abstract}
We are developing a general, unified, and rigorous analytical
framework for using gravitational lensing by compact objects to
test different theories of gravity beyond the weak-deflection limit.
In this paper we present the formalism for computing corrections
to lensing observables for static, spherically symmetric gravity
theories in which the corrections to the weak-deflection limit
can be expanded as a Taylor series in one parameter, namely the
gravitational radius of the lens object.  We take care to derive
coordinate-independent expressions and compute quantities that
are directly observable.  We compute series expansions for the
observables that are accurate to second-order in the ratio
$\vep = \vthbh/\vthE$ of the angle subtended by the lens's
gravitational radius to the weak-deflection Einstein radius,
which scales with mass as $\vep \propto \Mbh^{1/2}$.  The
positions, magnifications, and time delays of the individual
images have corrections at both first- and second-order in $\vep$,
as does the differential time delay between the two images.
Interestingly, we find that the first-order corrections to the total
magnification and centroid position vanish in all gravity theories
that agree with general relativity in the weak-deflection limit,
but they can remain nonzero in modified theories that disagree
with general relativity in the weak-deflection limit.  For the
\RN\ metric and a related metric from heterotic string theory,
our formalism reveals an intriguing connection between lensing
observables and the condition for having a naked singularity,
which could provide an observational method for testing the
existence of such objects.  We apply our formalism to the Galactic
black hole and predict that the corrections to the image positions
are at the level of 10 micro-arcseconds, while the correction to
the time delay is a few hundredths of a second.  These corrections
would be measurable today if a pulsar were found to be lensed by
the Galactic black hole; and they should be readily detectable
with planned missions like MAXIM.
\end{abstract}


\maketitle

\section{Introduction}
\label{sec:intro}

The gravitational deflection of light provided one of the first
observational tests of general relativity.  Now it is routinely
observed in a broad array of astrophysical contexts ranging from
stars through galaxies and clusters of galaxies up to the large-scale
structure of the universe.  (See \cite{schneider,petters} for
thorough discussions of gravitational lensing, and \cite{Saas-Fee}
for a recent review of astrophysical applications.)  All of the
effects seen so far occur in the weak-deflection, quasi-Newtonian
limit of general relativity.  Thus, while gravitational lensing has
proven valuable for measuring masses in astrophysics, it has not yet
been able to test fundamental theories of gravity in physics.

There has been significant theoretical effort over several decades
to understand lensing in the strong-deflection regime (e.g.,
\cite{darwin,atkinson,luminet, ohanian,virbhadra,frittelli,perlick,
bozza-strong,eiroa-RN,petters}), in particular for situations in
which (i) the lens is compact, static, and spherically symmetric,
(ii) the metric is asymptotically flat far from the lens, and
(iii) the source and observer lie in the asymptotically flat regime.
(See \cite{perlick} for an approach that does not require
asymptotic flatness.)  The resulting theory has yielded a remarkable
prediction: There should be an infinite series of images very close
to and on either side of the black hole's photon sphere, corresponding
to light rays that loop around the black hole once, twice, etc.\
before traveling to the observer (e.g., \cite{darwin,atkinson,luminet,
ohanian,virbhadra,bozza-strong}).  The possibility that these
``relativistic images'' could be used to test strong-deflection
gravity has created considerable excitement, inspiring extensive
studies of their properties in various familiar metrics (e.g.,
\cite{darwin,atkinson,luminet,ohanian,virbhadra,bozza-strong,eiroa-RN})
as well as those arising from string theory and braneworld gravity
\cite{bhadra,string-brane-strong,majumdar}, together with assessments
of prospects for detecting the images \cite{petters-SgrA,bozza-SgrA}.

Unfortunately, the relativistic images are exceedingly faint (a flux
correction of order $10^{-14}$ for the Galactic black hole
\cite{virbhadra,petters-SgrA}), which makes it important to consider
whether there are any other observable effects that can be used to
test theories of gravity.  The primary and secondary lensed images
--- which travel from the source to observer without looping around
the lens --- do not usually pass close enough to the lens's photon
sphere to experience extreme strong-deflection lensing, but they
may nevertheless be affected by various orders of
post-post-Newtonian (PPN) correction terms.  Since these images are
much easier to detect, it is valuable to compute the corrections to
their observable properties.  The first-order corrections have been
studied for the Schwarzschild metric (which is standard point-mass
lensing \cite{schneider,petters}), the \RN\ metric \cite{eiroa-RN},
and metrics with general PPN terms \cite{PPN-2,ebina} and mass
currents \cite{sereno}.  Certain aspects of higher-order PPN
corrections have been studied \cite{sarmiento}.  The lowest-order
(i.e., weak-deflection) theory has been studied for metrics from
string theory and braneworld gravity \cite{bhadra,kar,majumdar}.
Weyl gravity has been investigated extensively \cite{pireaux} and
so will not be treated here.  Unfortunately, much of the existing
work seems to comprise a diverse collection of results, some of them
rediscovered several times, that makes it difficult to draw general
conclusions.  Also, most of the previous studies have focused on
the light bending angle rather than on quantities that can be
observed directly in extra-solar lensing scenarios.

Our goal is to develop a general framework for computing corrections
to a core set of observable properties of the primary and secondary
lensed images in a general geometric theory of gravity.  In the
process, we shall unify the previous lensing analyses into a common
framework, and extend them to a higher order of approximation.  At
the same time, we shall demonstrate how to handle general gravity
theories using PPN terms up to third-order.  One crucial part of our
formalism is that we take care to work with {\em observable} quantities
and avoid coordinate dependence.  For instance, most previous studies
of corrections to the weak-deflection bending angle have expressed
the results in terms of the light ray's radial distance of closest
approach to the lens.  However, such a radial distance is a
coordinate-dependent quantity.  Ambiguities created by different
choices of coordinates can be alleviated by working instead with the
impact parameter of the light ray \cite{bangle-coordindep}.  The
impact parameter is an invariant of the light ray (a constant of
motion), and is given geometrically by the perpendicular distance,
relative to initial observers at infinity, from the center of the
lens to the asymptotic tangent line to the light ray trajectory at
the observer.  It is the quantity to use when defining the observable
angular position of a lensed image and will play a key role in our
formalism.

There are several reasons to compute corrections to a higher order
than has been done before.  One is that certain lensing observables
have first-order corrections that vanish in general relativity
(see, e.g., \cite{ebina}).  Our higher-order formalism will allow
us to understand the fortuitous cancellations that cause those
terms to vanish, and to find the lowest-order non-vanishing
corrections.  By being very general, we will also determine whether
the cancellations are generic or restricted to specific families of
gravity theories; we shall show, for example, that the cancellations
also occur in all theories of gravity that agree with general
relativity in the weak-deflection limit.  A second reason for
working to high order is that carrying the expansion far enough
will, in principle, allow us to bridge the gap that now exists
between weak- and strong-deflection analyses of lensing by black
holes and compact objects.  

Our generality will also allow us to find an unexpected connection
between lensing observables and certain kinds of naked singularities.
The \RN\ metric is usually taken to describe a charged black hole
in general relativity, but if the charge parameter exceeds a
threshold value, then the metric describes a naked singularity
instead.  We shall discover that the corrections to certain lensing
observables can have a negative value only if the condition for a naked
singularity is satisfied.  This could provide a direct observational
test for these exotic singularities, which are conventionally ruled
out by the still-unproved Cosmic Censorship conjecture
\cite{CosmicCensorship}.

This is the first in a series of papers intended to develop the
complete lensing framework.  Here we begin with a thorough analysis
of lensing by a static, spherically symmetric compact body, using a
formalism that can handle all gravity theories in which the metric
can be expressed as a series expansion in the single parameter
$\gravr$ (the gravitational radius of the compact body).  In
separate papers we will generalize to metrics with two parameters
and metrics that describe a rotating compact body.  In addition to
presenting the formalism, we will discuss some possible
astrophysical applications.  It is our hope that these studies will
be relevant to the next generation of black hole imagers, such as
MAXIM \cite{MAXIM}.

\section{Basic Assumptions of the Formalism}
\label{sec:formalism}

Since we seek to lay out a general framework, we should begin by
stating the assumptions very clearly.  Consider a compact body of
mass $\Mbh$, perhaps a black hole or neutron star, that is
described by a geometric theory of gravity.  This means that the
body's gravitational field is determined by a spacetime metric
appropriate to the theory of gravity in question.  (See
\S \ref{sec:metric} for the form of the metric.)  Possibilities
include general relativity, whose spacetime metric obeys the
Einstein equation, and modified gravity theories, so named
because their metrics are not governed by Einstein's equation.

\begin{figure}
\includegraphics[width=3.2in]{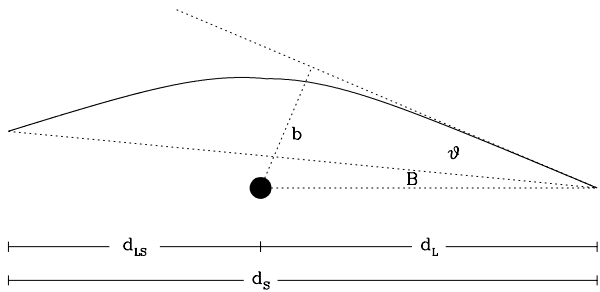}
\caption{
Schematic diagram of the lensing geometry.
}\label{fig:geom}
\end{figure}

We study how the body acts as a gravitational lens by considering
light rays that travel directly from the source to the observer
without looping around the compact body.  \reffig{geom} gives a
schematic diagram of the lensing situation and defines standard
quantities:
$\cb$ is the angular position of the unlensed source;
$\vth$ is the angular position of an image;
$\hat{\alpha}$ is the bending angle of the light ray;
and $d_L$, $d_S$, and $d_{LS}$ are the observer-lens,
observer-source, and lens-source angular diameter distances,
respectively.  Also important is the impact parameter $b$, which
is an invariant of the light ray (a constant of motion) and is
given geometrically by the perpendicular distance, relative to
inertial observers at infinity, from the center of the compact
body to the asymptotic tangent line to the light ray trajectory
at the observer.  As noted in \S \ref{sec:intro}, when defining
observables it is crucial to use the invariant impact parameter
rather than the coordinate-dependent distance of closest approach
(also see \cite{bangle-coordindep}).  The angular image position
is defined in terms of the impact parameter as $\vth = \sin^{-1}(b/d_L)$.

From the figure, elementary trigonometry establishes the
relationship (see \cite{virbhadra})
\beq \label{eq-lebh-1}
  \tan \cb = \tan \vth - D \ (\tan \vth + \tan (\hat{\alpha} - \vth))\,.
\eeq
where $D = d_{LS}/d_S$.  This equation agrees very well with the
full relativistic formalism for light propagation
\cite{frittelli}.  In this paper,
we take it
as the general form of the gravitational lens equation.

Henceforth, the angles describing image positions are assumed to
be positive, which then forces the source's angular position to take
on a positive or negative value depending on the image's location.  
Explicitly, the source is assumed to be fixed at an angle $\cb$,
which is taken to be positive when studying an image on the same
side of the compact object as the source, and negative when studying
an image on the opposite side.  (Note the different conventions for
the signs of $\vth$ and $\cb$.)

We make the following assumptions:
\begin{itemize}

\item[\underline{\bf A1:}] The gravitational lens is compact,
static, and spherically symmetric, with an asymptotically flat
spacetime geometry sufficiently far from the lens.  The spacetime
is vacuum outside the lens and flat in the absence of the lens.

\item[\underline{\bf A2:}] The observer and source lie in the
asymptotically flat regime of the spacetime.

\item[\underline{\bf A3:}] The light ray's distance of closest
approach $r_0$ and impact parameter $b$ both lie well outside the
gravitational radius $\gravr = G \Mbh/c^2$, namely,
$\gravr/r_0 \ll 1$ and $\gravr/b \ll 1$.  The bending angle can
then be expressed as a series expansion as follows in the single
quantity $\gravr/b$:
\beq \label{eq-bangle-ser}
  \hat{\alpha}(b) = A_1 \left(\frac{\gravr}{b}\right)
    \ + \ A_2 \left(\frac{\gravr}{b}\right)^2
    \ + \ A_3 \left(\frac{\gravr}{b}\right)^3
    \ + \ \order{\frac{\gravr}{b}}{4} .
\eeq
The coefficients $A_i$ are independent of $\gravr/b$, but may
include other fixed parameters of the spacetime.  Since $b$ and
$\gravr$ are invariants of the light ray, eq.~(\ref{eq-bangle-ser})
is independent of coordinates.  Note that the subscript of $A_i$
conveniently indicates that the component is affiliated with a
term of order $i$ in $\gravr/b$.

\end{itemize}

Assumption A1 is a natural first step to take for the study of
lensing by neutron stars and black holes.  The spherical symmetry
conveniently allows us to restrict attention to light rays moving
in a plane.  As we shall see in \S \ref{sec:applications},
assumption A2 certainly holds in interesting astrophysical settings.

As for assumption A3, the form of the general bending angle
$\hat{\alpha}$ in eq.~(\ref{eq-bangle-ser}) is a series expansion
expressing corrections to the standard weak-deflection bending angle
in general relativity,
\beq
  \hat{\alpha}_{\rm wf} (\vth) = 4 \left(\frac{\gravr}{b}\right)
  \approx 4 \left(\frac{\vthbh}{\vth}\right) ,
\eeq
where $\vthbh = \tan^{-1}(\gravr/d_L)$ is the angle subtended by
the gravitational radius of the compact object, and ``$\approx$''
simply indicates that in the last expression we have used the
standard small-angle approximation.  (This approximation is not
valid when we work beyond linear order, as discussed in
\S \ref{sec:lenseqn}.)  By writing the cofficient of the linear
term as $A_1$, we allow for the possibility that gravity theories
can differ slightly from general relativity in the quasi-Newtonian
regime.  In practice, though, observational analyses imply that
$A_1$ is quite close to 4:  for example, recent compilations of
20 years' worth of data imply that $A_1 = 3.999 66 \pm 0.000 90$
\cite{sh}.  There are as yet no good observational constraints on
the higher-order coefficients $A_2$ and $A_3$ in (\ref{eq-bangle-ser}),
so they are to be determined from the gravity theory in question
(as discussed below).

\section{Light Bending Angle in Various Gravity Theories}
\label{sec:bendangle}

A key part of any lensing framework is the light bending angle.
We state the general form of the metric we consider, and derive
the exact, general expression for the bending angle.  We obtain
the desired series expansion of the bending angle for the
Schwarzschild metric, working to higher order than has been done
previously, and taking care to express the result in terms of the
invariant impact parameter.  We then repeat the analysis for a
general post-post-Newtonian (PPN) metric, and illustrate how
various gravity theories can be studied with this approach.

\subsection{Form of the Metric and Bending Angle}
\label{sec:metric}

From assumption A1, the spacetime geometry is postulated to be
static, spatially spherically symmetric, and asymptotically
Minkowski:
\beq
  {\rm d}s^2 = - \bar{A}(\bar{r})\,{\rm d}t^2
    + \bar{B}(\bar{r})\,{\rm d}\bar{r}^2 + \bar{C}(\bar{r})\,{\rm d}\Omega^2\,,
\eeq
where ${\rm d} \Omega^2$ is the standard unit sphere metric, and
$\bar{A}(\bar{r}) \rightarrow 1$,
$\bar{B}(\bar{r}) \rightarrow 1$,
$\bar{C}(\bar{r}) \rightarrow \bar{r}^2$
as $\bar{r} \rightarrow \infty$.
Again by A1, the metric is Minkowski in the absence of the lens
and we make the natural mathematical assumptions (for regularity
in coordinate changes and constancy of causal structure) that
$\bar{A}$, $\bar{B}$, $\bar{C}$, $d\bar{A}/d\bar{r}$, and
$d\bar{C}/d\bar{r}$ are all positive in the region outside the
lens through which the light rays of interest propagate.
Note that we do not require that the metric approaches the GR
weak-deflection limit for $\bar{r}$ sufficiently large.  Since
$d\bar{C}/d\bar{r} >0$, the function $\bar{C}(\bar{r})$ is
invertible and allows a new radial coordinate
$r = \sqrt{\bar{C}(\bar{r})}$.
This transforms the metric as follows:
\beq \label{eq-metric}
  {\rm d}s^2 = - A(r)\,{\rm d}t^2 + B(r)\,{\rm d}r^2 + r^2\,{\rm d}\Omega^2\,.
\eeq

Because of the spherical symmetry, the geodesics of (\ref{eq-metric})
lie in a plane, which we can take to be the equatorial plane.  It
then suffices to work with a metric of the form
\beq \label{eq-metric-equatorial}
  {\rm d}s^2 = - A(r)\,{\rm d}t^2
    \ + \ B(r)\,{\rm d} r^2
    \ + \ r^2\, {\rm d}\varphi^2\,,
\eeq
where $\varphi$ is the azimuthal angle.  The Lagrangian $f$
determining the null geodesics of the metric
(\ref{eq-metric-equatorial}) is given by
\beq
f(r,v_t,v_r,v_\varphi) = \frac{1}{2} \left[- A(r)\, v_t^2
        \ + \ B(r)\, v_r^2 \ + \ r^2 \, v_\varphi^2\right]\,,
\eeq
where $(v_t,v_r,v_\varphi)$ is a velocity vector with components
along the $\partial_t$, $\partial_r$, and $\partial_\varphi$
coordinate directions, respectively.  Note that this Lagrangian
is independent of $t$ and $\varphi$.  The stationary curves are
governed by the Euler-Lagrange equations, which yield the geodesic
equations and provide two constants $E$ and $L$ of their motion:
\beq
  \frac{\partial f}{\partial \dot{t}} = E\,, \qquad
  \frac{\partial f}{\partial \dot{\varphi}} = L\,,
\eeq
which give
\beq \label{eq-E-L}
  \dot{t}(s) = \frac{E}{A(r(s))}\ , \qquad
  \dot{\varphi}(s) = \frac{L}{r^2(s)}\ .
\eeq
Here, a dot denotes differentiation with respect to an affine
parameter $s$ along the geodesic.  In addition, restricting the
Lagrangian to the position-velocity path of the null geodesic (in
the tangent bundle over spacetime) gives a third constant of the
motion:
\beqa
  0 &=& f(r(s),\dot{t}(s),\dot{r}(s),\dot{\varphi}(s)) \\
  &=& - A(r(s))\, \dot{t}^2(s) \ + \ B(r(s))\, \dot{r}^2(s)
     \ + \ r^2(s) \, \dot{\varphi}^2(s)\,. \nonumber
\eeqa
This yields
\beq \label{eq-rdot}
  \dot{r} = \pm \, |L| \, \sqrt{\frac{1/b^2 \ - \ A/r^2}{AB} }\ ,
\eeq
where $b = |L/E|$ and is called the {\em impact parameter}.
If the light ray's distance $r_0$ of closest approach to the lens
occurs at affine parameter $s = s_0$ such that $\dot{r}(s_0)=0$,
then $b$ and $r_0$ are simply related by
\beq \label{eq-genb}
  \frac{1}{b^2} = \frac{A (r_0)}{r_0^2}.
\eeq

Equations (\ref{eq-E-L}) and (\ref{eq-rdot}) then yield
\beq \label{eq-dvarphi-dr}
  \frac{d \varphi}{dr} = \pm \, \frac{1}{r^2} \
  \sqrt{\frac{AB}{1/b^2 \ - \ A/r^2}}
\eeq
(since $C > 0$ and $d \varphi/dr = \dot{\varphi}/\dot{r} =
\pm |\dot{\varphi}/\dot{r}|$).  The plus (minus) sign corresponds
to the portion of the light's trajectory where $r$ increases
(decreases) as a function of $\varphi$.  Similarly, equations
(\ref{eq-E-L}) and (\ref{eq-rdot}) also give
\beq \label{eq-dt-dr}
  \frac{d t}{dr} = \pm \, \frac{1}{b A} \
  \sqrt{\frac{AB}{1/b^2 \ - \ A/r^2}},
\eeq
where the plus (minus) sign corresponds to the portion of the
orbit where $r$ increases (decreases) as a function of $t$.

Consider a light ray that originates in the asymptotically
flat region of spacetime and is deflected by the compact body
before arriving at an observer in the flat region.  Equation
(\ref{eq-dvarphi-dr}) then yields the following expression for
the bending angle (e.g., \cite{weinberg}):
\beqa
  \hat{\alpha}(r_0) &=&
    2 \int_{r_0}^{\infty} \left|\frac{d \varphi}{ d r}\right| {\rm d} r
    \ - \ \pi \nonumber\\
  &=& 2 \int_{r_0}^{\infty} \frac{1}{r^2}\,
      \sqrt{\frac{AB}{1/b^2 \ - \ A/r^2}}\ {\rm d} r \ - \ \pi\,.
\label{eq-genal}
\eeqa

\subsection{General Relativity: Schwarzschild Metric}
\label{sec:GR-Schwarzschild}

\subsubsection{Schwarzschild Metric}

For a spherical, electrically neutral compact body, general
relativity yields a unique metric that is asymptotically flat
and obeys the vacuum Einstein equation: the Schwarzschild metric.
It is characterized by the single parameter $\gravr$ and can be
written in the form of (\ref{eq-metric}) with
\beq \label{eq-Sch-metric}
  A(r) =  1 - \frac{2\gravr}{r}\ ,
\qquad
  B(r) = \left(1 - \frac{2\gravr}{r}\right)^{-1} .
\eeq

We need to relate the distance of closest approach $r_0$ to the
impact parameter $b$.  Using eq.~(\ref{eq-genb}), we can write
$b$ in terms of $r_0$:
\beq \label{eq-rb}
  \frac{r_0}{b} = \sqrt{1 - 2 \frac{\gravr}{r_0}}\ .
\eeq
Inverting this to find $r_0$ in terms of $b$ yields
(cf.\ \cite{wald}, p. 145)
\beq \label{eq-rsolution-b}
  \frac{r_0}{b} = \frac{2}{\sqrt{3}}\  
  \cos\left[\frac{1}{3} \, \cos^{-1} 
\left(- \frac{3^{3/2} \ \gravr}{b} \right) \right]
\eeq

\begin{widetext}

\noindent
Since $\gravr/b \ll 1$ by A3, we can Taylor expand the right-hand
side in powers of $\gravr/b$ to obtain
\beq
  r_0 = b \,\Biggl[  1 - \frac{\gravr}{b}
    - \frac{3}{2}     \left( \frac{\gravr}{b}\right)^2
    - 4               \left( \frac{\gravr}{b}\right)^3
    - \frac{105}{8}   \left( \frac{\gravr}{b}\right)^4
    - 48              \left( \frac{\gravr}{b}\right)^5
    - \frac{3003}{16} \left( \frac{\gravr}{b}\right)^6
    + \order{\frac{\gravr}{b}}{7} \Biggr] .
\label{eq-rb-ser}
\eeq

\subsubsection{Schwarzschild Bending Angle}

From eqs.~(\ref{eq-genal}) and (\ref{eq-Sch-metric}), the exact
form of the light bending angle can be written as
\beq
  \hat{\alpha}(r_0) = 2 \int_{r_0}^\infty
    \frac{dw}{w^2 \sqrt{1/b^2 - 1/w^2 + 2 \gravr/w^3}} \ - \ \pi.
\eeq
We change variables to $x = r_0/w$, and also substitute for the
impact parameter $b$ using (\ref{eq-rb}).  This yields
\beq
  \hat{\alpha}(r_0) = 2 \int_0^1
    \frac{dx}{\sqrt{1 - 2 h - x^2 + 2 h\, x^3}} \ - \ \pi\,,
\eeq
where $h = \gravr/r_0$.  (Note that for computational purposes,
a numerically stable expression can be obtained by further changing
variables to $x = \cos\eta$.)  For pedagogical purposes, rewrite
the integral as 
\beq \label{eq-bangle-integral}
 \int_0^1 \frac{dx}{\sqrt{1 - 2 h -  x^2 + 2 h\, x^3}} = 
  \int_0^1 \frac{dx}{ \sqrt{1 - x^2}\ \sqrt{1 - 2 h (1 - x^3)/(1 - x^2)} }\ .
\eeq
By assumption A3 we have $h < 1/3$, which means physically that
the light ray is outside the photon sphere at $3 \gravr$.  The
rational function $(1 - x^3)/(1 - x^2)$ is monotonically increasing
on $[0,1]$ with a maximum value of $3/2$ at $x = 1$.   This means
that $0\le 2 h (1 - x^3)/(1 - x^2) < 1$, so we can Taylor expand
$[1 - 2 h (1 - x^3)/(1 - x^2)]^{-1/2}$ in a geometric series.
Carrying out the integration in eq.~(\ref{eq-bangle-integral})
term by term then gives
\beqa
  \hat{\alpha}(h) &=& 4 h + \left(-4 + \frac{15}{4} \pi\right) h^2
   \ + \ \left(\frac{122}{3} - \frac{15}{2} \pi\right) h^3
   \ + \ \left(-130 + \frac{3465}{64} \pi\right) h^4
   \ + \ \left(\frac{7783}{10} - \frac{3465}{16} \pi\right) h^5 \nonumber\\
&& \ + \ \left(-\frac{21397}{6} + \frac{310695}{256} \pi\right) h^6
   \ + \ \order{h}{7}\,. \label{eq-bangle-h}
\eeqa
To obtain the bending angle in terms of the invariant impact
parameter $b$, we use eq.~(\ref{eq-rb-ser}) to write $r_0$ in
terms of $b$.  This yields
\beq \label{eq-bangle-b}
  \hat{\alpha}(b) = A_1 \left(\frac{\gravr}{b}\right)
   + A_2 \left(\frac{\gravr}{b}\right)^2
   + A_3 \left(\frac{\gravr}{b}\right)^3
   + A_4 \left(\frac{\gravr}{b}\right)^4
   + A_5 \left(\frac{\gravr}{b}\right)^5
   + A_6 \left(\frac{\gravr}{b}\right)^6
   + \order{\frac{\gravr}{b}}{7} ,
\eeq
where
\beq
  A_1 = 4\,, \quad
  A_2 = \frac{15\pi}{4}\ , \quad
  A_3 = \frac{128}{3}\ , \quad
  A_4 =  \frac{3465 \pi}{64}\ , \quad
  A_5 = \frac{3584}{5}\ , \quad
  A_6 = \frac{255255 \pi}{256}\ .
\eeq
It follows that eq.~(\ref{eq-bangle-b}) has the form required by
assumption A3.  We shall not actually use terms beyond
$\order{\gravr/b}{3}$, but we have included some higher order
terms because they have not appeared in the literature before.

\subsection{PPN Approach}
\label{sec:PPN}

\subsubsection{PPN Metric to Third Order}

The post-post-Newtonian (PPN) formalism is a convenient way to
handle the wide range of gravity theories in which the weak-deflection
limit can be expressed as a series expansion in the single variable
$\gravr$.  The formalism was extended to third-order in
\cite{sarmiento}.  However,  we reformulate that treatment to have
a better fit with our approach, and present new results.  

Express the coefficients of the standard metric (\ref{eq-metric})
in a PPN series to third-order as follows: 
\beqa
A(r) &=& 1 \ + \ 2\, a_1 \, \left(\frac{\phi}{c^2}\right)
  \ + \ 2\, a_2\, \left(\frac{\phi}{c^2}\right)^2
  \ + \ 2\, a_3\, \left(\frac{\phi}{c^2}\right)^3 
  \ + \ \ldots\,, \label{eq-PPN-A} \\
B(r) &=& 1 \ - \ 2\, b_1 \, \left(\frac{\phi}{c^2}\right)
  \ + \ 4\, b_2\, \left(\frac{\phi}{c^2}\right)^2
  \ - \ 8\, b_3\, \left(\frac{\phi}{c^2}\right)^3
  \ + \ \ldots\,, \label{eq-PPN-B}
\eeqa
where $\phi$ is the three-dimensional Newtonian potential with
\beq
\frac{\phi}{c^2} = - \frac{\gravr}{r'}
\eeq

If the metric is in isotropic form, namely,
\beq \label{eq-metric-isotropic}
  {\rm d}s^2 = - A'(r')\,{\rm d}t^2
    \ + \ B'(r')\, [ {\rm d}r'^2 + r'^2 \, {\rm d}\Omega^2]\,,
\eeq
then the PPN convention is to write
\beqa
A'(r') &=& 1 \ + \ 2\, \alpha' \, \frac{\phi'}{c^2}
  \ + \ 2\, \beta' \,\left(\frac{\phi'}{c^2}\right)^2
  \ + \ \frac{3}{2}\, \xi' \,\left(\frac{\phi'}{c^2}\right)^3
  \ + \ \ldots\,, \\
B'(r') &=& 1 \ - \ 2\, \gamma' \, \frac{\phi'}{c^2} 
  \ + \ \frac{3}{2}\, \delta' \, \left(\frac{\phi'}{c^2}\right)^2
  \ - \ \frac{1}{2}\, \eta' \, \left(\frac{\phi'}{c^2}\right)^3
  \ + \ \ldots\,,
\eeqa
where $\phi'/c^2 = -\gravr/r'^2$, 
and $(\alpha', \beta', \gamma', \delta', \xi', \eta')$ denote
the Eddington-Robertson parameters (with primes added to avoid
confusion with standard lensing quantities).  The parameters are
chosen so that the Schwarzschild metric has
\beq
  \alpha' = \beta' = \gamma'= \delta' = \xi' = \eta' = 1\,.
\eeq

We can relate the parameters in the standard and isotropic forms
of the metric by comparing the time, radial, and angular parts of
the two metrics, yielding the relations
\beq
  A(r) = A'(r')\,, \qquad
  r^2 = B'(r')\, r'^2\,, \qquad
  B(r)\,{\rm d}r^2 = B'(r')\,{\rm d}r'^2\,.
  \label{eq-PPNreln}
\eeq
The second and third relations yield
\beq
  \ln r' = \int \frac{\sqrt{B(r)}}{r}\,dr + \mbox{const}\,,
\eeq
where the constant is chosen so that $r'/r \to 1$ as $r \to \infty$.
Plugging $r'$ into (\ref{eq-PPNreln}) and identifying terms on the
left- and right-hand sides that have the same order in $\gravr/r$,
we find the following correspondence between the standard and
isotropic coefficients:
\beqa
  a_1 &=& \alpha'\,, \\
  b_1 &=& \gamma'\,, \\
  a_2 &=& \beta' - \alpha' \gamma'\,, \\
  b_2 &=& \frac{3\delta' + \gamma'^2}{4}\ , \\
  a_3 &=& \frac{3\xi' + 3\alpha' \delta' - 8 \beta' \gamma'
    + 2 \alpha' \gamma'^2}{4}\ , 
 \label{eq-a3}\\
  b_3 &=& \frac{3\eta' + 15 \delta' \gamma' - 2 \gamma'^3}{16}
 \label{eq-a4}\ .
\eeqa
The correspondences (\ref{eq-a3}) and
(\ref{eq-a4})  have not been worked out before in the
literature.  For reference, the Schwarzschild metric has
\beq
  a_1 = b_1 = b_2 = b_3 = 1\,, \qquad
  a_2 = a_3 = 0\,.
\eeq

We need to relate the distance of closest approach $r_0$ to the
invariant impact parameter $b$.  Using eq.~(\ref{eq-genb}), we can
write $b$ in terms of $r_0$:
\beq \label{eq-PPN-bofr} 
  b = r_0 \Biggl[ 1 + a_1 \frac{\gravr}{r_0}
    + \frac{3a_1^2-2a_2}{2} \left(\frac{\gravr}{r_0}\right)^2
    + \frac{5a_1^3-6a_1 a_2+2a_3}{2} \left(\frac{\gravr}{r_0}\right)^3
    + \order{\frac{\gravr}{r_0}}{4} \Biggr] .
\eeq
To invert this relation and find $r_0$ in terms of $b$, we postulate
a relation of the form
\beq
  r_0 = b \left[ 1 + c_1 \frac{\gravr}{b}
    + c_2 \left(\frac{\gravr}{b}\right)^2
    + c_3 \left(\frac{\gravr}{b}\right)^3
    + \order{\frac{\gravr}{b}}{4} \right] ,
\eeq
plug this into (\ref{eq-PPN-bofr}), and solve for the constants $c_i$
by requiring that the coefficient of each power of $\gravr$ vanishes.
This yields
\beq \label{eq-PPN-rofb}
  r_0 = b \Biggl[ 1 - a_1 \frac{\gravr}{b}
    + \frac{2a_2-3a_1^2}{2} \left(\frac{\gravr}{b}\right)^2
    - \left(a_3 - 4 a_1 a_2 + 4 a_1^3\right) \left(\frac{\gravr}{b}\right)^3
    + \order{\frac{\gravr}{b}}{4} \Biggr] .
\eeq
The result (\ref{eq-PPN-rofb}) is also new.

\subsubsection{PPN Bending Angle}

To compute the light bending angle, we take the exact expression
(\ref{eq-genal}), plug in the PPN metric functions 
(\ref{eq-PPN-A})--(\ref{eq-PPN-B}), and change integration variables
to $x=r_0/w$.  We also substitute for the impact parameter $b$ using
eq.~(\ref{eq-PPN-bofr}).  Finally, we introduce $h = \gravr/r_0$, so
that the series expansion becomes a Taylor series in $h$.  

The first step yields
\beqa
  \hat{\alpha}(h) &=& 2 \int_0^1 \frac{dx}{\sqrt{x^2-1}} \Biggl\{
1 + h \left(b_1 x + \frac{a_1}{1+x}\right)
\nonumber\\
&&\qquad
+ \ h^2 \Biggl[ - a_2 - \frac{1}{2}(b_1^2 - 4 b_2) x^2
  + \frac{3 a_1^2}{2(1+x)^2} +  (2 a_1^2 + a_1 b_1)\frac{x}{1+x} \Biggr]
\nonumber\\
&&\qquad
+ \ \frac{h^3}{2} \Biggl[ a_1 (8 a_1^2 + 4 a_1 b_1 - b_1^2 + 4 b_2) (x-1)
  + 4 a_1 a_2 (1-2 x) + 2 x (a_3 - a_2 b_1)
\nonumber\\
&&\qquad\qquad\quad
  + (b_1^3 - 4 b_1 b_2 + 8 b_3) x^3 + \frac{5 a_1^3}{(1+x)^3}
  - \frac{3 a_1^2 (4 a_1 + b_1)}{(1+x)^2} + \frac{2 a_3}{1+x}
\nonumber\\
&&\qquad\qquad\quad
  + \frac{a_1 (20 a_1^2 - 10 a_2 + 7 a_1 b_1 - b_1^2 + 4 b_2}{1+x} \Biggr]
  + \order{h}{4} \Biggr\} \ - \ \pi\,.
\eeqa
Carrying out the integration term by term then yields
\beqa
  \hat{\alpha}(h) &=& 2(a_1+b_1) h
\ + \ h^2 \left[ -2 a_1 (a_1+b_1) + \left( 2 a_1^2 - a_2 + a_1 b_1
   - \frac{b_1^2}{4} + b_2 \right) \pi \right] \nonumber\\
&&
+ \ h^3 \Biggl[ \frac{67}{3} a_1^3 - 18 a_1 a_2 + 4 a_3 + 9 a_1^2 b_1
  - 2 b_1 (a_2 + a_1 b_1) + 8 a_1 b_2 + \frac{2}{3} \Bigl( b_1^3 - 4 b_1 b_2
  + 8 b_3 \Bigr) \nonumber\\
&&\label{eq-alpha-h-PPN}
  \qquad\quad
   - a_1 \Bigl( 4 a_1^2 - 2 a_2 + 2 a_1 b_1 - \frac{b_1^2}{2}
   + 2 b_2 \Bigr) \pi \Biggr]
+ \order{h}{4}, 
\eeqa
which agrees with \cite{sarmiento}.  However, the expression
(\ref{eq-alpha-h-PPN}) is coordinate-dependent since it is
written in terms of the distance of closest approach.  To obtain
an invariant expression (so we can discuss observable quantities),
we use (\ref{eq-PPN-rofb}) to write $r_0$ in terms of $b$ and
obtain:
\beq
  \hat{\alpha}(b) = A_1 \left(\frac{\gravr}{b}\right)
    \ + \ A_2 \left(\frac{\gravr}{b}\right)^2
    \ + \ A_3 \left(\frac{\gravr}{b}\right)^3
    \ + \ \order{\frac{\gravr}{b}}{4} ,
\eeq
where
\beqa
  A_1 &=& 2(a_1+b_1)\,,
\label{eq-PPN-A1}\\
  A_2 &=& \left( 2 a_1^2 - a_2 + a_1 b_1 - \frac{b_1^2}{4} + b_2 \right) \pi\,,
\label{eq-PPN-A2}\\
  A_3 &=& \frac{2}{3} \Bigl[ 35 a_1^3 + 15 a_1^2 b_1
    - 3 a_1 \left(10 a_2 + b_1^2 - 4 b_2 \right)
    + 6 a_3 + b_1^3 - 6 a_2 b_1 - 4 b_1 b_2 + 8 b_3 \Bigr] .
\label{eq-PPN-A3}
\eeqa
This invariant bending angle expression has not appeared in the
literature before.

\subsection{Sample Gravity Theories}
\label{sec:sample}

In general relativity, the \RN\ metric describes a black hole with 
physical charge $Q$.  Although charge is generally expected to
become neutralized, in certain circumstances it could persist for
1,000 to 10,000 years \cite{punsly}, perhaps allowing detection of
such black holes.  The \RN\ metric has the form of (\ref{eq-metric})
with the metric functions
\beq \label{eq-RNmetric}
  A(r) = B(r)^{-1} = 1 - \frac{2\gravr}{r} + \frac{q^2\,\gravr^2}{r^2}\ ,
\eeq
where $q = \sqrt{G}\, Q/(c^2 \gravr)$ is a dimensionless parameter.  
If $q^2 > 1$, the metric actually describes a naked singularity
rather than a charged black hole (e.g., \cite{eiroa-RN,bozza-strong}).
It is natural to view (\ref{eq-RNmetric}) as a series expansion in
$\gravr/r$.  We can then identify the PPN coefficients:
\beq
  a_1 = 1\,, \quad
  a_2 = \frac{q^2}{2}\,, \quad
  a_3 = 0\,, \quad
  b_1 = 1\,, \quad
  b_2 = 1 - \frac{q^2}{4}\,, \quad
  b_3 = 1 - \frac{q^2}{2}\,. \nonumber
\eeq
Using eqs.~(\ref{eq-PPN-A1})--(\ref{eq-PPN-A3}), we can determine
the coefficients in the expansion of the bending angle:
\beq
  A_1 = 4\,, \quad
  A_2 = (5-q^2) \frac{3\pi}{4}\ , \quad
  A_3 = \frac{128}{3} - 16 q^2\,.
\eeq
Notice that if $q \to 0$ we recover the results for the Schwarschild
metric.  Also notice that the condition for a naked singularity
corresponds to $A_2 < 3\pi$.  Turning this around, we may say that
{\em for the \RN\ metric of general relativity, the coefficient
$A_2$ can be negative only if there is a naked singularity}.

In heterotic string theory, modifications of the Einstein equation
lead to a different charged black hole solution, which is often called
the Gibbons-Maeda-Garfinkle-Horowitz-Strominger (GMGHS) black hole
\cite{gibbons}.  In \cite{bhadra}, the metric's lensing properties
were computed to lowest order in both the weak- and strong-deflection
regimes, although only with a coordinate-dependent approach.  We can
include the weak-deflection limit in our invariant formalism, and
thereby obtain new results.  For a black hole with gravitational
radius $\gravr$ and charge $Q$, the GMGHS metric is often written in
the form
\beq
  {\rm d}s^2 = - \left( 1 - \frac{2 \gravr}{\bar{r}} \right)\,{\rm d}t^2
  + \left( 1 - \frac{2 \gravr}{\bar{r}} \right)^{-1} {\rm d}\bar{r}^2
  + \bar{r}^2 \left( 1 - \frac{q^2 \gravr}{\bar{r}} \right) {\rm d} \Omega^2\,,
\eeq
where again $q = \sqrt{G} Q/(c^2 \gravr)$.  As in the \RN\ case,
this metric describes a naked singularity if the charge parameter
exceeds some threshold, in this case $q^2 > 2$.  We convert to
standard coordinates by setting
\beq
  r^2 = \bar{r}^2 \left( 1 - \frac{q^2 \gravr}{\bar{r}} \right) .
\eeq
This puts the metric into the form of (\ref{eq-metric}) with the
metric functions
\beqa
  A(r) &=& 1 - \frac{4 \gravr}{q^2 \gravr + \sqrt{q^4 \gravr^2 + 4 r^2}}\ , \\
  B(r) &=& \left(1 - \frac{4 \gravr}{q^2 \gravr + \sqrt{q^4 \gravr^2 + 4 r^2}}
    \right)^{-1} \frac{4 r^2}{\gravr^2 q^4 + 4 r^2}\ . \nonumber\\
\eeqa
Expanding the metric functions as Taylor series in $\gravr/r$, we
can identify the PPN parameters
\beq
  a_1 = 1\,, \quad
  a_2 = \frac{q^2}{2}\,, \quad
  a_3 = \frac{q^4}{8}\,, \quad
  b_1 = 1\,, \quad
  b_2 = 1 - \frac{q^2}{4} - \frac{q^4}{16}\ , \quad
  b_3 = 1 - \frac{q^2}{2} - \frac{q^4}{32}\ .
\eeq
The coefficients in the expansion of the bending angle are then
\beq
  A_1 = 4\,, \qquad
  A_2 = \left(60 - 12 q^2 - q^4\right) \frac{\pi}{16}\ , \qquad
  A_3 = \frac{128}{3} - 16 q^2\,.
\eeq
In this case, the condition for a naked singularity corresponds to
$A_2 < 2\pi$.  Or, we may again say that {\em for the GMGHS metric
$A_2$ can be negative only if there is a naked singularity.}

\section{Lensing Framework in Various Gravity Theories}
\label{sec:lensing}

We can now move beyond the bending angle to compute corrections
to the observable properties of the primary and secondary lensed
images.  In this section we focus on the positions and magnifications
of the images; time delays are deferred to \S \ref{sec:Tdel}.

\subsection{Lens Equation}
\label{sec:lenseqn}

We start with the general lens equation,
\beq
  \tan \cb = \tan \vth - D \ (\tan \vth + \tan (\hat{\alpha} - \vth)) ,
\eeq
and seek an appropriate series expansion.  First, we change
variables to match the scalings commonly used in the astrophysical
lensing literature.  A natural scale is the weak-deflection angular
Einstein ring radius,
\beq
  \vthE = \sqrt{\frac{4 G \Mbh d_{LS}}{c^2 d_L d_S}}\ .
\eeq
We then define:
\beq \label{eq-newvar}
  \beta = \frac{\cb}{\vthE}\ , \quad
  \theta = \frac{\vth}{\vthE}\ , \quad
  \vep = \frac{\vthbh}{\vthE} = \frac{\vthE}{4\,D}\ .
\eeq
In other words, $\beta$ and $\theta$ are the scaled angular positions
of the source and image, respectively.  The quantity $\vep$ represents
the angle subtended by the gravitational radius normalized by the
angular Einstein radius, and it becomes our new expansion parameter.

The second step is to postulate that the solution of the lens
equation can be written as a series expansion of the form
\beq \label{eq-tseries}
  \theta = \theta_0 + \theta_1\,\vep + \theta_2\,\vep^2 + 
    \order{\vep}{3} ,
\eeq
where $\theta_0$ is expected to be the image position in the
weak-deflection limit, and the coefficients $\theta_1$ and $\theta_2$
of the correction terms remain to be determined.  (This is a
standard perturbation theory analysis; e.g., \cite{perttheory}.)

After making these substitutions, we first find that we can write
the bending angle as
\beq
  \hat{\alpha} = \frac{A_1}{\theta_0}\,\vep
   \ + \ \frac{A_2 - A_1\theta_1}{\theta_0^2}\,\vep^2
   \ + \ \frac{1}{\theta_0^3} \left[ A_3 - 2 A_2 \theta_1
         + A_1 \left( \frac{8}{3} D^2 \theta_0^4
         + \theta_1^2 - \theta_0 \theta_2 \right)
         \right] \vep^3
   \ + \ \order{\vep}{4} .
\eeq
Note that since we are expanding beyond linear order, it is
important to use the exact geometric relations between physical
and angular radii: $\vth = \sin^{-1}(b/d_L)$ and
$\vthbh = \tan^{-1}(\gravr/d_L)$.  The standard small angle
approximations ($\vth \approx b/d_L$ and
$\vthbh \approx \gravr/d_L$) are valid only at linear order.

Now making the substitutions in the lens equation and Taylor
expanding in $\vep$, we find:
\beqa
  0 &=& D \left[ -4\beta + 4\theta_0 - \frac{A_1}{\theta_0} \right] \vep
\ + \ \frac{D}{\theta_0^2} \Biggl[ -A_2
    + \left( A_1+4\theta_0^2 \right) \theta_1 \Biggr]\,\vep^2
\label{eq-lebh-2}\\
&&
- \frac{D}{3\theta_0^3} \Biggl\{ A_1^3 + 3 A_3 - 12 A_1^2 D \theta_0^2
    + A_1 (56 D^2 \theta_0^4 + 3 \theta_1^2 - 3 \theta_0 \theta_2)
    - 2 \Bigl[ 32 D^2 \theta_0^3 (\theta_0^3-\beta^3) + 3 A_2 \theta_1
    + 6\theta_0^3 \theta_2 \Bigr] \Biggr\} \vep^3 
+ \order{\vep}{4} \nonumber
\eeqa
This is the desired series expansion of the lens equation.

\subsection{Image Positions}
\label{sec:imgpos}

We now solve eq.~(\ref{eq-lebh-2}) term by term to find the
coefficients $\theta_i$ in the series expansion for the image
position.  The idea is to fix the source position $\beta$ and
find the values of $\theta_i$ that make each term in
(\ref{eq-lebh-2}) vanish.  The first-order term is just the
standard weak-deflection lens equation,
\beq \label{eq-lens0}
  0 = -\beta + \theta_0 - \frac{1}{\theta_0}\ ,
\eeq
which yields the weak-deflection image position
\beq
\label{eq-theta0}
  \theta_0 = \frac{1}{2} \left( \beta + \sqrt{\beta^2+4} \right) .
\eeq
We neglect the negative solution because we have explicitly
specified that angles describing image positions are positive.
We find the positive-parity image $\theta_0^+$, which lies on the
same side of the lens as the source, by using a positive angular
source position, $\beta > 0$.  We then find the negative-parity image
$\theta_0^-$, which lies on the opposite side of the lens from the
source, by using $\beta < 0$.  In other words, we can rewrite
(\ref{eq-theta0}) as
\beq \label{eq-theta0p}
  \theta_0^\pm = \frac{1}{2}\left(\sqrt{ 4 + \beta^2} \pm |\beta| \right) .
\eeq
One curious feature of the positive-
and negative-parity image positions in the weak-deflection limit
is that
\beq \label{eq-tptm=1}
  \theta_0^{+}\ \theta_0^{-} = 1\,.
\eeq
Note that (\ref{eq-theta0p}) makes it clear that (\ref{eq-tptm=1})
does not depend on the sign of $\beta$.

Next, we choose $\theta_1$ to make the second-order term in
(\ref{eq-lebh-2}) vanish.  This yields
\beq \label{eq-theta1}
  \theta_1 = \frac{A_2}{A_1+4\theta_0^2}\ .
\eeq
For the third-order term, we still have $\theta_0$ given by
(\ref{eq-theta0}) and $\theta_1$ given by (\ref{eq-theta1}), and
we must choose $\theta_2$ to make the coefficient of $\vep^3$ in
(\ref{eq-lebh-2}) vanish.  This yields
\beqa
  \theta_2= &=& \frac{1}{3\,\theta_0\,(A_1+4\theta_0^2)^3} \Biggl\{
A_1 (A_1^4-3 A_2^2+3 A_1 A_3-64 A_1 D^2) \nonumber\\
&&\quad
\ - \ 4 \Bigl[6 A_2^2 - 2 A_1 (A_1^3 + 3 A_3) + 3 A_1^4 D
    - 16 A_1 D^2 (3 A_1 - 8)\Bigr] \theta_0^2 \nonumber\\
&&\quad
\ + \ 8 \Bigl[2 A_1^3(1-6D) + 6 A_3 - 128 D^2
    + A_1 D^2 (192+A_1(-24+7A_1))\Bigr] \theta_0^4 \nonumber\\
&&\quad
\ + \ 64 D \Bigl[48 D - 24 A_1 D + A_1^2 (-3+7 D)\Bigr] \theta_0^6
\ + \ 128 D^2 \Bigl(-24+7A_1\Bigr) \theta_0^8 \Biggr\} . \label{eq-theta2}
\eeqa
Note that we have used the relation $\beta = \theta_0 - 1/\theta_0$
to replace $\beta$ on the right-hand side.  It is possible to rewrite
the right-hand side in terms of $\beta$ instead of $\theta_0$, but
at this point we believe that is less useful.

For gravity theories with $A_1 = 4$ (including general relativity)
the correction terms simplify slightly to
\beqa
  \theta_1 &=& \frac{A_2}{4(1+\theta_0^2)}\ , \\
  \theta_2 &=& \frac{1}{48\,\theta_0\,(1+\theta_0^2)^3} \Biggl[
    -3 A_2^2 (1+2\theta_0^2) + \ 4 (1+\theta_0^2)^2 \Bigl( 3 A_3 + 64
    - 64 D^2 - 192 D \theta_0^2
    + 192 D^2 \theta_0^2 + 32 D^2 \theta_0^4 \Bigr) \Biggr] .
\eeqa
It is worth pointing out that the image position can be written to
first-order in $\vep$ as
\beq
  \theta = \theta_0 \ + \ \frac{A_2}{A_1+4\theta_0^2}\ \vep
    \ + \ \order{\vep}{2} .
\eeq
The sign of the first-order correction to the weak-deflection
image position is given by the sign of $A_2$ (for both the
positive- and negative-parity images).  We have already seen
that the sign of $A_2$ is connected with the possible presence
of naked singularities: in the \RN\ and GMGHS metrics, $A_2$
can be negative only if there is a naked singularity (see
\S \ref{sec:sample}).  Thus, the correction to the image
positions can provide a possible observational test for naked
singularities.

\subsection{Magnifications}
\label{sec:mag}

The signed magnification $\mu$ of a lensed image at angular position
$\vth$ is
\beq
  \mu(\vth) = \left[\frac{\sin \cb (\vth)}{\sin \vth} \ 
    \frac{d \cb(\vth)}{d \vth} \right]^{-1} .
\eeq
We change to our scaled variables as in eqs.~(\ref{eq-newvar}) and
(\ref{eq-tseries}), make a Taylor expansion in $\vep$, and then
substitute for $\theta_1$ and $\theta_2$ using (\ref{eq-theta1})
and (\ref{eq-theta2}).  In this way we obtain a series expansion
for the magnification:
\beq \label{eq-muser}
  \mu = \mu_0 \ + \ \mu_1\,\vep \ + \ \mu_2\,\vep^2
    \ + \ \order{\vep}{3} ,
\eeq
where
\beqa
  \mu_0 &=& \frac{16\theta_0^4}{16\theta_0^4-A_1^2}\ , \\
  \mu_1 &=& - \frac{16 A_2 \theta_0^3}{(A_1+4\theta_0^2)^3}\ , \\
  \mu_2 &=& \frac{8\theta_0^2}{3 (A_1-4\theta_0^2)^2 (A_1+4\theta_0^2)^5}
    \Biggl\{ A_1^4 D^2 (512 - 9 A_1^3)
\ + \ 4 A_1^3 \theta_0^2 \Bigl[ -384 A_1 D^2 + A_1^3 (4+12D-9D^2)
    + 4 (3 A_3 + 256 D^2) \Bigr] \nonumber\\
&&\quad
+ \ 64 A_1 \theta_0^4 \Bigl[ A_1^4(1+3D) - 9 A_2^2 + 3 A_1 A_3
    + 128 A_1 D^2 - 192 A_1^2 D^2 + 24 A_1^3 D^2 \Bigr] \nonumber\\
&&\quad
- \ 256\,\theta_0^6 \Bigl[ -9 A_2^2 + 3 A_1 A_3 + 96 A_1^2 D^2 - 48 A_1^3 D^2
    + A_1^4 (1+3D+2D^2) \Bigr] \nonumber\\
&&\quad
- \ 256\,\theta_0^8 \Bigl[ 12 A_3 - 96 A_1^2 D^2 + A_1^3 (4+12D+7D^2) \Bigr]
\ + \ 1024 A_1^2 D^2 \theta_0^{10} \Biggr\} . \label{eq-mu2}
\eeqa
Recall that the sign of the signed magnification indicates the
parity of the image: $\mu > 0$ for $\theta^+$, the positive-parity
primary image; while $\mu < 0$ for $\theta^-$, the negative-parity
secondary image. 

For a gravity theory with $A_1 =4$, these expressions reduce to
\beqa
  \mu_0 &=& \frac{\theta_0^4}{\theta_0^4-1}\ , \\
  \mu_1 &=& - \frac{A_2 \theta_0^3}{4(1+\theta_0^2)^3}\ , \\
  \mu_2 &=& - \frac{\theta_0^2}{24 (\theta_0^2-1) (\theta_0^2+1)^5} \Biggl\{
768 D \theta_0^2 (\theta_0^2+1)^2 
    - 64 D^2 (\theta_0^2+1)^2 \Bigl(1 + 16 \theta_0^2 + \theta_0^4\Bigr) \\
&&\qquad\qquad\qquad\qquad\qquad
+ \theta_0^2 \Bigl[ 256 + (512 - 9 A_2^2) \theta_0^2 + 256 \theta_0^4
    + 12 A_3 (\theta_0^2+1)^2 \Bigr] \Biggr\} . \nonumber
\eeqa

It is worth pointing out that if $A_1=4$ the first-order changes
in the magnifications of the positive- and negative-parity images
have the following relation:
\beq \label{eq-mu1pm}
  \mu_1^{+} \ = \ -\ \frac{A_2\,(\theta_0^{+})^3}{4[1+(\theta_0^{+})^2]^3}
% \ = \ -\ \frac{A_2\,(\theta_0^{-})^{-3}}{4[1+(\theta_0^{-})^{-2}]^3}
  \ = \ -\ \frac{A_2\,(\theta_0^{-})^3}{4[(\theta_0^{-})^2+1]^3}
  \ = \ \mu_1^{-}\,.
\eeq
In the second equality we used (\ref{eq-tptm=1}) to write
$\theta_0^{+} = 1/\theta_0^{-}$.  In gravity theories with
$A_1 = 4$ and $A_2 > 0$, the $\mu_1$ perturbation is negative,
so it makes $\mu^+$ less positive (fainter) and $\mu^-$ more
negative (brighter) by exactly the same amount.  (The opposite
occurs for $A_1 = 4$ and $A_2 < 0$.)  In other words, the
magnifications of the positive- and negative-parity images are
shifted by the same amount but in the opposite sense.  This has
important implications for the total magnification and centroid
(see below).

Notice that the sign of the first-order magnification correction
depends on the sign of $A_2$, so this provides another possible
observational test for naked singularities.

\subsection{Total Magnification and Centroid}
\label{sec:totmag}

If the two images are too close together to be resolved (as in
microlensing), the main observables are the total magnification
and the magnification-weighted centroid position.  Using our
results above, we find the total magnification
$\mu_{\rm tot} = |\mu^+| + |\mu^-|$ to be
\beq
  \mu_{\rm tot} = \frac{16 A_1^2 (\theta_0^8-1)}
    {(16 \theta_0^4 - A_1^2) (A_1^2 \theta_0^4-16)}
\ - \ \frac{16 (A_1-4) A_2 \theta_0^3}{(A_1+4\theta_0^2)^3 (4+A_1\theta_0^2)^3}
  \Bigl\{ [16+A_1(4+A_1)](\theta_0^6-1) + 12 A_1 \theta_0^2 (\theta_0^2-1)
  \Bigr\}\,\vep 
\ + \ \order{\vep}{2} . \label{eq-magtot}
\eeq
The second-order term can be worked out from (\ref{eq-mu2}) if
desired, but is too complicated to write here.  Notice that the sign
of $A_2$ again determines the sign of the first-order correction.

It is important to see the factor of $(A_1-4)$ multiplying the
first-order term in (\ref{eq-magtot}).  It means that {\em the
first-order correction to the total magnification vanishes with
full generality in any gravity theory with $A_1=4$.}  In such
theories, the first-order changes in the magnifications of the
positive- and negative-parity images exactly cancel (see
eq.~\ref{eq-mu1pm}).  A correction then enters at second-order,
which is given by
\beq
 \mu_{\rm tot} = \frac{2 + \beta^2}{\beta \, \sqrt{4 + \beta^2}}
\ + \ \frac{1}{12\,\beta\,(4 + \beta^2)^{5/2}} \Biggl\{
9 A_2^2 - 12 A_3 (4+\beta^2) 
 - 64 (4+\beta^2) \left[ 4 + 12 D - (18+\beta^2) D^2 \right] \Bigr\}\,\vep^2
\ + \ \order{\vep}{3} . \label{eq-magtot-A1=4}
\eeq
Here we have found it convenient to express the result in terms
of the source angular position $\beta$.  

The magnification-weighted centroid position is defined by
\beq
  \Theta_{\rm cent} =  \frac{\theta^{+} |\mu^{+}| - \theta^{-} |\mu^{-}|}
    {|\mu^{+}| + |\mu^{-}|}
  = \frac{\theta^{+} \mu^{+} + \theta^{-} \mu^{-}}
    {\mu_{+} - \mu_{-}}\ .
\eeq
(The sign in the numerator may be understood by recalling that
we use positive angles, $\theta_{\pm} > 0$.)  Our perturbation
analysis then yields
\beqa
  \Theta_{\rm cent} &=& \frac{A_1^2 (\theta_0^8 - \theta_0^6 + \theta_0^4
    - \theta_0^2 + 1) - 16 \theta_0^4}{A_1^2 \theta_0 (\theta_0^6
    - \theta_0^4 + \theta_0^2 - 1)} \label{eq-centroid}\\
&&
+ \ \frac{A_2 (A_1-4) \theta_0^2}{A_1^4 (\theta_0^4-1) (\theta_0^4+1)^2
      (A_1+4\theta_0^2) (4+A_1\theta_0^2)} 
\Biggl[ A_1^2 (16+4A_1+A_1^2) (1+\theta_0^{12}) 
\ - \ 4 A_1 (32-A_1^2) \theta_0^2 (1+\theta_0^8) \nonumber\\
&&\qquad\quad
  + \ (256-64 A_1-32 A_1^2+8 A_1^3+A_1^4) \theta_0^4 (1+\theta_0^4)
\ + \ (256+32 A_1^2+24 A_1^3+A_1^4) \theta_0^6 \Biggr]\,\vep 
\ + \ \order{\vep}{2} . \nonumber
\eeqa
As before the second-order term can be worked out but is too
complicated to write here.  Notice again that the sign of $A_2$
yet again determines the sign of the first-order correction.

We again see a factor of $(A_1-4)$ multiplying the first-order term.
{\em Thus, the first-order correction to the centroid vanishes
exactly in any gravity theory with $A_1 = 4$.}  In such theories,
a correction appears at second-order and is given in terms of
$\beta$ by
\beqa
\Theta_{\rm cent} &=& \frac{\beta(3+\beta^2)}{2+\beta^2} 
  \ - \ \frac{\beta}{24 (4+\beta^2) (2+\beta^2)^2} \Biggl\{
  9 A_2^2 - 12 A_3 (4+\beta^2) - 128 (4+\beta^2) (2-D^2) \nonumber\\
&&\qquad\qquad\qquad\qquad
  - 64 (4+\beta^2) \left[ (9+\beta^2) D - 6 \right] D \beta^2
  \Biggr\}\,\vep^2 \ + \ \order{\vep}{3} .
\label{eq-centroid-A1=4}
\eeqa
Notice that in this case ($A_1=4$), the sign of $A_2$ does not
affect the corrections, but the sign of $A_3$ does.

\section{Time Delays in Various Gravity Theories}
\label{sec:Tdel}

Having studied the positions and brightnesses of the primary and
secondary images, we are now ready to compute their time delays.
The analysis parallels some of what has gone before, but is
different enough to warrant a separate treatment.

\subsection{General Formalism}

Let $R_{\rm src}$ and $R_{\rm obs}$ be the radial coordinates of
the source and observer, respectively.  From geometry relative
to the flat metric of the distant observer, who is assumed to be
at rest in the natural coordinates of the metric
(\ref{eq-metric-equatorial}), we can work out (see \reffig{geom})
\beq \label{eq-Rdef}
  R_{\rm obs} = d_L\,, \qquad
  R_{\rm src} = \left( d_{LS}^2 + d_S^2 \tan^2 \cb \right)^{1/2} .
\eeq
The radial distances are very nearly the same as angular diameter
distances since the source and observer are in the asymptotically
flat region of the spacetime.  In other words, the distortions in
distances near the black hole are assumed to have little impact on
the total flat metric distance from the compact body to the observer
or source.  We focus on spacetimes that would be flat in the absence
of the lens (see assumption A1), and in that case the light ray
would travel along a linear path from the source to the observer
with length $d_S/\cos\cb$.

The time delay is the difference between the light travel time for
the actual ray, and the travel time for the ray the light would have
taken had the lens been absent.  This can be written as
\beq \label{eq-tdel}
  c \tau = T(R_{\rm src}) + T(R_{\rm obs}) - \frac{d_S}{\cos\cb}\ ,
\eeq
with
\beq
  T(R) = \int_{r_0}^{R} \left| \frac{dt}{dr}\right| \, {\rm d} r
  = \frac{1}{b} \, \int_{r_0}^{R} \frac{1}{A(w)} \
    \sqrt{\frac{A(w) \, B(w) }{1/b^2 \ - \ A(w)/w^2}} \, {\rm d} w\,,
\eeq
where we have used eq.~(\ref{eq-dt-dr}) for $dt/dr$.  Unlike
eq.~(\ref{eq-genal}), this integral cannot extend to infinity
because the travel time would diverge.  

Note that there are no redshift factors in these equations because
we are assuming a spacetime that is static and asymptotically flat.
This assumption is not overly restrictive because expected
applications involve non-cosmological lens systems (see
\S \ref{sec:applications}).

\subsection{General Relativity: Schwarzschild Metric}

To compute $T(R)$ for the Schwarzschild metric, we take the metric
functions from eq.~(\ref{eq-Sch-metric}), change integration
variables to $x = r_0/w$, and use (\ref{eq-rb}) to relate $b$ to
$r_0$, so that we find

\beq
  T(R) = r_0 \int_{r_0/R}^{1} x^{-2} \left( 1 - 2\,h\,x \right)^{-1}
    \left( 1 - x^2\ \frac{1-2\,h\,x}{1-2\,h} \right)^{-1/2}\ dx\,,
\eeq
where $h = \gravr/r_0$.  Expanding the integrand as a Taylor series
in $h$ and integrating term by term yields
\beqa
  T(R) &=& \sqrt{R^2-r_0^2} \ + \ h\,r_0\,\left[ \frac{\sqrt{1-\xi^2}}{1+\xi}
    + 2 \ln\left(\frac{1+\sqrt{1-\xi^2}}{\xi}\right) \right]
\label{eq-Sch-Tser} \\
  && + \ h^2\,r_0\,\left[
    \frac{15}{2} \left( \frac{\pi}{2} - \sin^{-1}\xi \right)
    - \frac{(4+5\xi)\sqrt{1-\xi^2}}{2(1+\xi)^2} \right] \nonumber\\
  && + \ h^3\,r_0\,\left[
    - \frac{15}{2} \left( \frac{\pi}{2} - \sin^{-1}\xi \right)
    + \frac{(60+157\xi+133\xi^2+35\xi^3)\sqrt{1-\xi^2}}{2(1+\xi)^3} \right]
  \ + \ \order{h}{4} , \nonumber
\eeqa
where $\xi = r_0/R$.

This expression is currently written in terms of the distance of
closest approach $r_0$, but we want to rewrite it in terms of the
invariant impact parameter $b$.  We can use (\ref{eq-rb-ser}) to
make the translation, but we must then consider the nature of the
series expansion.  The expansion in $h=\gravr/r_0$ naturally becomes
an expansion in $\gravr/b$.  It may be less obvious, but we also
want to expand in $b/R$, which is small because both the source and
observer lie far from the lens.  To do the joint expansion, we need
to consider the amplitudes of $\gravr/b$ and $b/R$.  In terms of
angular variables, we have two fairly simple cases,
\beqa
  \frac{\gravr}{b} &\sim& \frac{\vthbh}{\vthE} \sim \vep\,, \\
  \frac{b}{R_{\rm obs}} = \frac{b}{d_L} &\sim& \vthE \sim D\,\vep\,,
\eeqa
and one that is slightly more involved,
\beq
  \frac{b}{R_{\rm src}} = \frac{b}{\sqrt{d_{LS}^2 + d_S^2 \tan^2 \cb}}
    = \frac{1}{\sqrt{D^2 + \tan^2 \cb}}\ \frac{b}{d_S}
    = \frac{1-D}{\sqrt{D^2 + \tan^2 \cb}}\ \frac{b}{d_L}
    \sim \frac{D(1-D)}{\sqrt{D^2 + \tan^2 \cb}}\ \vep\,.
\eeq
The bottom line is that $\gravr/b$, $b/R_{\rm obs}$, and
$b/R_{\rm src}$ are all similar in amplitude, up to factors that
depend on $D$.  We will do the rigorous series expansion of the
time delay momentarily, but for now it is instructive to expand
(\ref{eq-Sch-Tser}) in both $\gravr/b$ and $b/R$, taking care
to collect terms of a given order in any combination of the two
quantities.  Working to third-order, we find:
\beq
  \frac{T(R)}{R} = 1
  \ - \ \left\{ \frac{1}{2}\,\frac{b}{R} \left[
        \frac{b}{R} - 2\,\frac{\gravr}{b}
        + 4\,\frac{\gravr}{b} \ln\left(\frac{1}{2}\,\frac{b}{R}\right) \right]
        \right\}
  \ + \ \left\{ \frac{15\pi}{4}\,\frac{b}{R}\,\left(\frac{\gravr}{b}\right)^2
        \right\}
  \ + \ \ldots\ 
  \label{eq-Sch-ToverR}
\eeq
There is no term that is linear in $\gravr/b$ or $b/R$.  The
first term in braces is of second-order, while the second is of
third-order.  We have neglected terms of fourth-order and higher.

That was meant to be pedagogical.  To properly compute the series
expansion for the full time delay (\ref{eq-tdel}), we first compute
$T(R_{\rm src})$ and $T(R_{\rm obs})$ by plugging (\ref{eq-Rdef})
into (\ref{eq-Sch-Tser}).  We replace $r_0$ with $b$ using
(\ref{eq-rb-ser}).  We change to angular variables using
$b = d_L \sin\vth$, and then reintroduce the scaled angular
variables $\theta$ and $\beta$ defined in (\ref{eq-newvar}).
Finally, we take a formal Taylor series in our expansion parameter
$\vep$.  Putting the pieces together, we find:
\beq
  c\tau = 8\,\frac{d_L\,d_{LS}}{d_S}\,\Biggl\{
    \left[ 1 + \beta^2 - \theta_0^2 - \ln \left(
    \frac{d_L\,\theta_0^2\,\vthE^2}{ 4\,d_{LS}} \right) \right] \vep^2
  \ + \
    \frac{15\pi - 8(1+\theta_0^2)\theta_1}{4\,\theta_0}\ \vep^3
    \ + \ \order{\vep}{4}
    \Biggr\} , \label{eq-Sch-tau1}
\eeq
The interpretation of this result becomes more clear when we
recognize that a characteristic scale for the time delay is
\beq \label{eq-tauE}
  \tau_E \equiv \frac{d_L d_S}{c\,d_{LS}}\,\vthE^2
  = 4\,\frac{\gravr}{c}\ .
\eeq
Although it is not obvious from the definition, the second equality
shows that $\tau_E$ is independent of the distances.  Using this
definition, we can write (\ref{eq-Sch-tau1}) as
\beq
  \frac{\tau}{\tau_E} =
    \frac{1}{2} \left[ 1 + \beta^2 - \theta_0^2 - \ln \left(
    \frac{d_L\,\theta_0^2\,\vthE^2}{ 4\,d_{LS}} \right) \right]
  \ + \ \frac{15 \pi - 8 (1+\theta_0^2) \theta_1}{8\,\theta_0}\ \vep
  \ + \ \order{\vep}{2} . \label{eq-Sch-tau2}
\eeq

Now rewrite $\theta_1$ (the first-order correction to the image
position) using (\ref{eq-theta1}), with $A_1 = 4$ and $A_2 = 15\pi/4$
for the Schwarzschild metric, to obtain
\beq
  \frac{\tau}{\tau_E} =
    \frac{1}{2} \left[ 1 + \beta^2 - \theta_0^2 - \ln \left(
    \frac{d_L\,\theta_0^2\,\vthE^2}{ 4\,d_{LS}} \right) \right]
  \ + \ \frac{15\pi}{16\,\theta_0}\ \vep
  \ + \ \order{\vep}{2} . \label{eq-Sch-tau3}
\eeq
Though not yet apparent, the zeroth-order term in (\ref{eq-Sch-tau3})
reduces to the familiar lensing time delay in the weak-deflection limit
of general relativity.  Rearranging, we can write the zeroth-order term
as
\beq
  \tau = \frac{d_L\,d_S}{c\,d_{LS}}
    \left[ \frac{(\vth_0 - \cb)^2}{2} - \vthE^2 \ln\vth_0 \right]
    \ + \ C\,,
\eeq
where we have used the identity $\beta = \theta_0 - \theta_0^{-1}$
from (\ref{eq-lens0}), and we have also used the definitions
$\theta_0 = \vth_0/\vthE$ and $\beta = \cb/\vthE$ from
(\ref{eq-newvar}).  The ``contant'' term $C$ in this expression
is independent of $\vth_0$ and $\cb$; it depends on $\vthE$ and
the distances as $C = \tau_E [1+\ln(d_L/4 d_{LS})]/2$.

Finally, to make contact with conventional calculations we examine
the differential time delay $\Delta\tau = \tau_{-} - \tau_{+}$
between the positive-parity primary image and the negative-parity
secondary image.  This can be written as
\beq \label{eq-Sch-dtau}
  \Delta\tau = \Delta\tau_0 \ + \ \vep\,\Delta\tau_1 \ + \ \order{\vep}{2} ,
\eeq
where
\beqa
  \Delta\tau_0 &=& \tau_E
  \left[ \frac{(\theta_0^{-})^{-2} - (\theta_0^{+})^{-2}}{2}
    - \ln\left(\frac{\theta_0^{-}}{\theta_0^{+}}\right) \right] , \\
  \Delta\tau_1 &=& \tau_E\ \frac{15\pi}{16}\,\frac{(\theta_0^{+}-\theta_0^{-})}
    {\theta_0^{+} \theta_0^{-}}\ .
\eeqa

It is possible to derive the second-order correction terms in
eqs.~(\ref{eq-Sch-tau3}) and (\ref{eq-Sch-dtau}), but they are more
complicated.  Since we have already found the first correction terms
for both absolute and differential time delays, it seems unnecessary
to write down the higher-order terms.

\subsection{Time Delay via the PPN Metric}

To compute $T(R)$ for the PPN approach, we take the metric functions
from eqs.~(\ref{eq-PPN-A}) and (\ref{eq-PPN-B}), use (\ref{eq-PPN-bofr})
to relate $b$ to $r_0$, change integration variables to $x=r_0/w$,
and finally carry out the integration to find
\beqa
  T(R) &=& \sqrt{R^2-r_0^2} \ + \ h\,r_0\,\left[
    a_1 \frac{\sqrt{1-\xi^2}}{1+\xi}
    + (a_1+b_1) \ln\left(\frac{1+\sqrt{1-\xi^2}}{\xi}\right) \right] \\
  && + \ h^2\,r_0\,\Biggl[
    \left( 4 a_1^2 - 2 a_2 + 2 a_1 b_1 - \frac{1}{2} b_1^2 + 2 b_2 \right)
      \left( \frac{\pi}{2} - \sin^{-1}\xi \right) \nonumber\\
  &&\qquad\qquad
    - a_1 \left( a_1 + b_1 + \frac{3}{2} a_1 \xi + b_1 \xi \right)
      \frac{\sqrt{1-\xi^2}}{2(1+\xi)^2} \Biggr] \nonumber\\
  && + \ h^3\,r_0\,\left[ F(\xi)\,\frac{\sqrt{1-\xi^2}}{2(1+\xi)^2}
    - a_1 \left( 4 a_1^2 - 2 a_2 + 2 a_1 b_1 - \frac{1}{2} b_1^2 + 2 b_2 \right)
      \left( \frac{\pi}{2} - \sin^{-1}\xi \right) \right] \nonumber\\
  && + \ \order{h}{4} , \nonumber
\eeqa
where $\xi = r_0/R$.  Here $F(\xi)$ is a cubic polynomial in $\xi$,
which is long and not particularly enlightening so we have not written
it out (but we do write complete expressions in what follows).

As in the Schwarzschild case, it is instructive to convert from the
distance of closest approach $r_0$ to the impact parameter $b$ (now
using eq.~\ref{eq-PPN-rofb}), and then expand in both $\gravr/b$ and
$b/R$ to find
\beq
  \frac{T(R)}{R} = 1
  \ - \ \left\{\frac{1}{2}\,\frac{b}{R} \left[
        \frac{b}{R} - 2 a_1\,\frac{\gravr}{b}
        + 2(a_1+b_1)\,\frac{\gravr}{b} \ln\left(\frac{1}{2}\,\frac{b}{R}\right)
    \right] \right\} 
  \ + \ \left\{ \left(8a_1^2 - 4 a_2 + 4 a_1 b_1 - b_1^2 + 4 b_2\right)
    \frac{\pi}{4}\,\frac{b}{R}\,\left(\frac{\gravr}{b}\right)^2 \right\}
  \ + \ \ldots\,, \label{PPN-ToverR}
\eeq
However, again we need to be more careful to obtain a rigorous
series expansion of the time delay.  Repeating the analysis
discussed around eqs.~(\ref{eq-Sch-tau1})--(\ref{eq-Sch-tau2}),
we obtain
\beq
  \frac{\tau}{\tau_E} =
    \frac{1}{2}
    \left[ a_1 + \beta^2 - \theta_0^2 - \frac{a_1+b_1}{2} \ln \left(
    \frac{d_L\,\theta_0^2\,\vthE^2}{ 4\,d_{LS}} \right) \right]
  \ + \ \frac{1}{8\,\theta_0}
    \Bigl[ (8 a_1^2 - 4 a_2 + 4 a_1 b_1 - b_1^2 + 4 b_2) \pi
    - 4 (a_1 + b_1 + 2 \theta_0^2) \theta_1 \Bigr] \vep
  \ + \ \order{\vep}{2} .
\label{eq-PPN-tau1}
\eeq
We can rewrite $\theta_1$ (the first-order correction to the
image position) using (\ref{eq-theta1}), with $A_1$ and $A_2$
given by (\ref{eq-PPN-A1}) and (\ref{eq-PPN-A2}), to obtain
\beq
  \frac{\tau}{\tau_E} =
    \frac{1}{2}
    \left[ a_1 + \beta^2 - \theta_0^2 - \frac{a_1+b_1}{2} \ln \left(
    \frac{d_L\,\theta_0^2\,\vthE^2}{ 4\,d_{LS}} \right) \right]
\ + \ \frac{\pi}{16\,\theta_0}
    \Bigl( 8 a_1^2 - 4 a_2 + 4 a_1 b_1 - b_1^2 + 4 b_2 \Bigr)\,\vep
\ + \ \order{\vep}{2} .  \label{eq-PPN-tau2}
\eeq
Finally, the differential time delay between the positive- and
negative-parity images is
\beq \label{eq-PPN-dtau}
  \Delta\tau = \Delta\tau_0 \ + \ \vep\,\Delta\tau_1 \ + \ \order{\vep}{2} ,
\eeq
where
\beqa
  \Delta\tau_0 &=& \tau_E
    \left[ \frac{(\theta_0^{-})^{-2} - (\theta_0^{+})^{-2}}{2}
    - \frac{a_1+b_1}{2}\,\ln\left(\frac{\theta_0^{-}}{\theta_0^{+}}\right)
    \right] , \\
  \Delta\tau_1 &=& \tau_E\ \frac{\pi}{16}
    \Bigl( 8 a_1^2 - 4 a_2 + 4 a_1 b_1 - b_1^2 + 4 b_2 \Bigr)
    \frac{(\theta_0^{+}-\theta_0^{-})}{\theta_0^{+} \theta_0^{-}}\ .
\eeqa

\end{widetext}

\section{Application to the Galactic Black Hole}
\label{sec:applications}

To illustrate our results, we consider gravitational lensing by the
supermassive black hole at the center of our Galaxy.  The black hole
has a mass of $\Mbh = (3.6\pm0.2) \times 10^{6}\,M_\odot$ \cite{ghez}
and is located at a distance of $d_L = 7.9 \pm 0.4$ kpc \cite{eisen}
from Earth.  (For illustration purposes, we adopt the nominal
values and neglect the small uncertainties.) Its gravitational
radius is therefore
$\gravr = 1.1 \times 10^{12}\mbox{ cm} = 3.4 \times 10^{-7}\mbox{ pc}$,
which corresponds to an angle of $\thbh = 9.0$ $\mu$as (micro-arcseconds).

Suppose the source lies a distance $d_{LS}$ beyond the black hole.
Typical distances will be $d_{LS} \sim 1$--100 pc, so we have
$d_L \approx d_S \gg d_{LS} \gg \gravr$ which means that both the
source and the observer lie in the asymptotically flat regime of the
spacetime, confirming assumption A2.  The angular Einstein radius is
$\vthE = 0.068\,(d_{LS}/10\mbox{ pc})^{1/2}$ as, which is much larger
than $\thbh$.  Since the primary and secondary images both lie near
the Einstein radius, each has a light ray with an impact parameter
and distance of closest approach that are much larger than the
gravitational radius, confirming that part of assumption A3.  Put
another way, our dimensionless expansion parameter is
\beq
  \vep = \frac{\thbh}{\vthE} = 1.3 \times 10^{-4} \times
    \left(\frac{d_{LS}}{10\mbox{ pc}}\right)^{-1/2} ,
\eeq
which is small enough to justify working with series expansions in
$\vep$.  The natural lensing time scale is $\tau_E = 71$ sec
(independent of the distances; see eq.~\ref{eq-tauE}).

\begin{figure}
\includegraphics[width=3.2in]{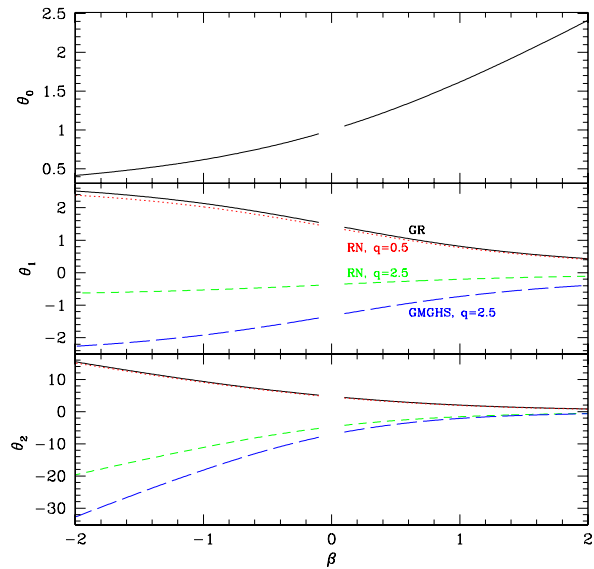}
\caption{
Terms in the series expansion (\ref{eq-tseries}) for the angular
image position, as a function of the angular source position.  Recall
that $\beta>0$ corresponds to the positive-parity image while
$\beta<0$ corresponds to the negative-parity image.  Sources very
close to the origin $|\beta|<0.1$ are not shown.
(Top) The zeroth-order image position $\theta_0$, which is the same
for all gravity theories.
(Middle) The first-order term $\theta_1$; recall that the full
correction term is $\vep\,\theta_1$.
Solid curve: Schwarzschild metric in general relativity.
Dotted and short-dashed curves: \RN\ metric in general relativity,
with charge parameter $q=0.5$ and 2.5, respectively.
Long-dashed curve: GMGHS metric in string theory (modified gravity),
with charge parameter $q=2.5$.  (The GMGHS case with $q=0.5$ is very
similar to the \RN\ case with $q=0.5$.)
(Bottom) The second-order term $\theta_2$.  Again, the full correction
term is $\vep^2 \theta_2$.
All angular lengths are in units of the angular Einstein radius
$\vthE$.  For the Galactic black hole, 
$\vthE = 0.068\,(d_{LS}/10\mbox{ pc})^{1/2}$ as and the dimensionless
expansion parameter is
$\vep = 1.3 \times 10^{-4} \times (d_{LS}/10\mbox{ pc})^{-1/2}$.  
}\label{fig:pos}
\end{figure}

We obviously want to compute the corrections to lensing observables
for the Schwarzschild metric in general relativity.  For comparison,
we also consider three other cases: a charged black hole described
by the \RN\ metric in general relativity, with charge parameter
$q=0.5$; and a naked singularity described by either the \RN\ metric
of general relativity or the GMGHS metric of heterotic string theory
(modified gravity), with ``charge'' parameter $q=2.5$.  (A charged
black hole described by the GMGHS metric with $q=0.5$ is very similar
to the \RN\ case with $q=0.5$, so we do not show it.)  Note that all
four cases are from two gravity theories in which the coefficient
of the leading term in the expansion of the bending angle is
$A_1 = 4$.

\begin{figure}
\includegraphics[width=3.2in]{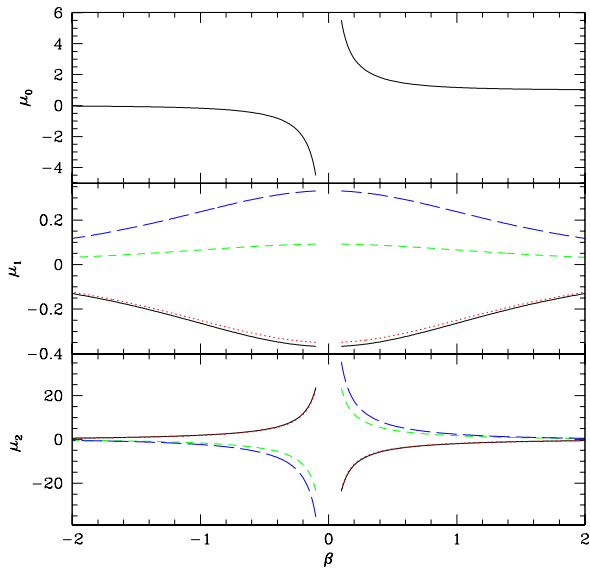}
\caption{
Similar to \reffig{pos}, but showing the terms in the series
expansion (\ref{eq-muser}) of the individual image magnifications.
Again recall that the full first- and second-order correction terms
are $\vep\,\mu_1$ and $\vep^2 \mu_2$, respectively.
}\label{fig:mag}
\end{figure}

\reffigs{pos}{mag} show the corrections to the individual image
positions and magnifications, as a function of the scaled angular
source position $\beta$, for the four sample cases.  The scaled
first-order correction term for the image position $\theta_1$
is of order unity, so the full correction to the image position
is of order $\vep\,\vthE \sim 10$ $\mu$as.  {\em The correction is
small, but detectable with high-resolution radio interferometry;
and it should be readily measurable with planned
microarcsecond-resolution missions such as {\rm MAXIM} } \cite{MAXIM}.
While the difference between neutral (Schwarzschild) and charged
(\RN) black hole cases is fairly small, the naked singularity case
stands out for having the opposite sign (see \S \ref{sec:imgpos}).
In other words, it will be challenging but not impossible to detect
the corrections to the usual weak-deflection lensing.  If the
corrections can be detected, high precision will be needed to
distinguish between different kinds of black holes.  However, it
will be easy to rule out certain kinds of naked singularities
observationally.

\begin{figure}
\includegraphics[width=3.2in]{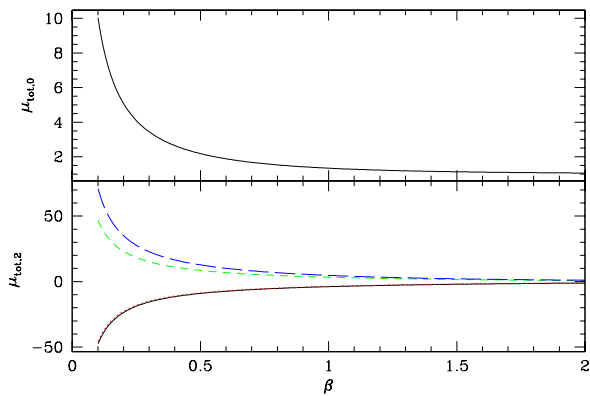}
\caption{
Zeroth- and second-order terms in the series expansion
(\ref{eq-magtot-A1=4}) of the total magnification.  Our examples
involve gravity theories with $A_1=4$, so the first-order correction
term vanishes identically and is not shown.  For each $\beta>0$, we
find the positive-parity image using the source position $\beta$,
and the negative-parity image using the source position $-\beta$,
and then combine them to obtain the total magnification.  The line
types are the same as in \reffig{pos}.
}\label{fig:mtot}
\end{figure}

\begin{figure}
\includegraphics[width=3.2in]{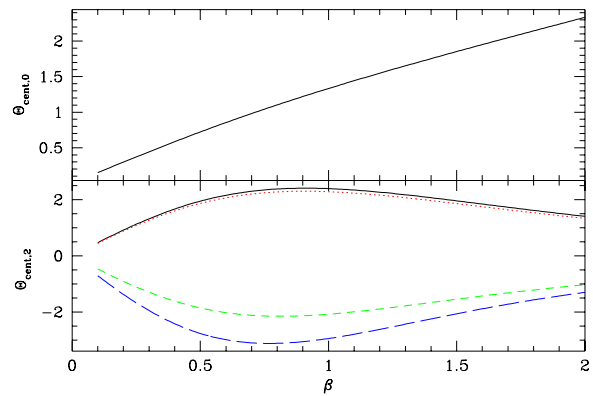}
\caption{
Similar to \reffig{mtot}, but showing the zeroth- and second-order
terms in the series expansion (\ref{eq-centroid-A1=4}) of the
magnification-weighted centroid position, in units of $\vthE$.
Again, for our sample gravity theories having $A_1=4$, the
first-order correction term vanishes identically and is not shown.
}\label{fig:cen}
\end{figure}

\reffigs{mtot}{cen} show the corrections to the total magnification
and magnification-weighted centroid position.  The corrections
to these observables are much smaller --- the centroid correction
is of order $\vep^2 \vthE \sim 10^{-9}$ as --- because the
first-order terms vanish in gravity theories with $A_1 = 4$.
Although interesting, these quantities will be very challenging
to use in the near future for realistic observational tests of
gravity theories.

\begin{figure}
\includegraphics[width=3.2in]{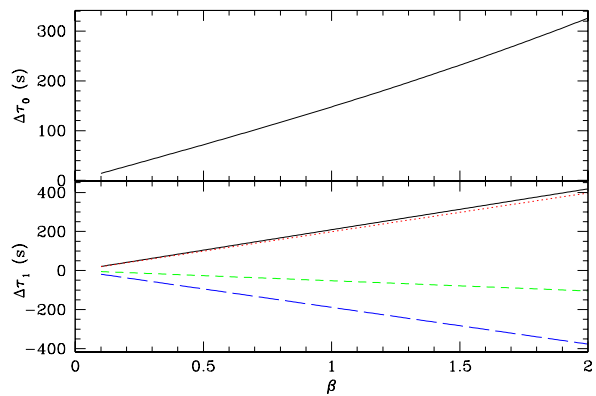}
\caption{
Zeroth- and first-order terms in the series expansion
(\ref{eq-PPN-dtau}) of the differential time delay between the two
images.  Note that we have defined $\Delta\tau_0$ and $\Delta\tau_1$
to have dimensions of time, so they are expressed in seconds.  Recall
that the full first-order correction is $\vep\,\Delta\tau_1$, where
the expansion parameter is
$\vep = 1.3 \times 10^{-4} \times (d_{LS}/10\mbox{ pc})^{-1/2}$
for the Galactic black hole.
}\label{fig:tau}
\end{figure}

Finally, \reffig{tau} shows the corrections to the differential
time delay between the positive- and negative-parity images.
{\em Remarkably, the first-order correction $\vep\,\Delta\tau_1$ 
is predicted to be as large as a few hundredths of a second ---
which would be relatively easy to measure if we could find a pulsar
lensed by the Galactic black hole.}  As with the image positions,
distinguishing between charged and neutral black holes would
require higher precision (not unfeasible in the case of a pulsar
source), but ruling out certain kinds of naked singularities would
be easy because they lead to corrections with the opposite sign.
Although we have not explicitly computed the second-order
corrections, we can estimate that they would be of order
$\vep^2 \tau_E \sim 10^{-6}$ s.

\section{Conclusions}
\label{sec:conclusions}

We have introduced a rigorous and general framework for computing
corrections to standard weak-deflection lensing observables.  In
this paper we have presented a formalism for handling any static,
spherically symmetric theory of gravity in which corrections to the
weak-deflection regime can be expanded as a Taylor series in the
gravitational radius of the compact body acting as a gravitational
lens.  Conceptually, three points distinguish our framework from
previous studies.  First, we take care to avoid coordinate dependence
by expressing our results in terms of invariant quantities.  Second,
we go beyond the bending angle to study quantities that are directly
observable for extra-solar lensing studies, including the positions,
magnifications, and time delays of the lensed images.  Third, our
general approach allows us to unify the diverse results that have
been presented before.

Besides the framework itself, our main results are series expansions
for the lensed image positions and magnifications that are accurate
to second-order in $\vep = \vthbh/\vthE$, or the ratio of the angle
subtended by the lens's gravitational radius to the (weak-deflection)
angular Einstein radius, as well as series expansions for the lensing
time delays with corrections at first-order in $\vep$.  The signs of
the first-order corrections are determined by the sign of the
coefficient $A_2$ in the series expansion of the light bending
angle.  If $A_2 > 0$, both the positive- and negative-parity images
are shifted away from the lens, and the positive-parity image gets
fainter while the negative-parity image gets brighter.  If $A_2 < 0$,
the opposite occurs.

The sign of $A_2$ appears to be connected with the possible
existence of naked singularities.  In the sample gravity theories
we have considered --- namely the Schwarzshild and \RN\ metrics in
general relativity, and the GMGHS metric in heterotic string theory
--- $A_2$ can be negative only if naked singularities exist.  This
connection between hypothetical naked singularities and observable
lensing quantities is exciting because it offers the possibility
that (certain kinds of) naked singularities can be ruled out
observationally rather than just by the still-unproved Cosmic
Censorship conjecture.  Note that we must use the qualifying
phrase ``certain kinds of'' because not all naked singularities
lead to $A_2 < 0$.  In these less extreme cases, we must go beyond
a gross feature like the sign of $A_2$ and consider the extent to
which fine, quantitative constraints on $A_2$ from realistic data
could rule out all or at least most interesting types of naked
singularities.

It has been known that for the Schwarzschild metric the first-order
corrections to the total magnification and magnification-weighted
centroid vanish (see \cite{ebina}).  We have shown that this result
is not generic: it depends on precise cancellations that can occur
only for gravity theories in which the leading-order term in the
series expansion of the bending angle has the value $A_1 = 4$.  In
practice, $A_1$ does equal 4 in the sample gravity theories we have
considered, and it is constrained by observational data to be quite
close to 4 (e.g.,\cite{sh}).  Nevertheless, it is important to
understand why these particular first-order corrections vanish, and
to recognize that it is possible to devise gravity theories for
which that is not the case.

We have applied our formalism to lensing by the Galactic black hole.
We predict the corrections to the image positions to be at the level
of 10 microarcseconds, and the correction to the time delay between
the images to be a few hundredths of a second.  The position
corrections would be measurable today with radio interferometry, if
we could find a radio source that is lensed by the Galactic black
hole.  The time delay correction would be measurable if the source
were a pulsar.  Even if such a convenient source cannot be found,
the corrections should be measurable with planned missions such as
MAXIM \cite{MAXIM}.

\begin{acknowledgments}

This work was supported by NSF grants DMS-0302812, AST-0434277,
and AST-0433809.  AOP would like to acknowledge the hospitality of
the American Institute of Mathematics for hosting a workshop on
Kerr black holes, where part of this work was conducted.  He 
enthusiastically thanks the participants, which included
researchers from Germany, Italy, Russian, and the U.S.A., for
informative presentations and discussions on the state of the art
in the field of black hole lensing.  Volker Perlick is also
acknowledged for emphasizing the importance of using the terms
``weak-deflection'' and ``strong-deflection'' as a precise way to
describe the various limits in black hole lensing.

\end{acknowledgments}

\end{document}